\documentclass[pra, twocolumn,  superscriptaddress, groupedaddress]{revtex4}
\usepackage{amssymb}
\usepackage{color}

\usepackage{amsthm}
\usepackage{graphicx}
\usepackage{dcolumn}
\usepackage{bm}
\usepackage{subfigure}
\usepackage{amssymb}
\usepackage{verbatim}
\usepackage{amscd}
\usepackage{amssymb}
\usepackage{amsmath, amsfonts}
\usepackage{setspace}
\usepackage{amsthm}
\usepackage{enumerate}

\usepackage{bbm}

\usepackage{subfigure}
\usepackage{varwidth,xcolor}

\usepackage[]{graphicx}

\newtheorem*{theorem*}{Theorem}
\newtheorem*{lemma*}{Lemma}

\begin{document}

\title{
Error correction for encoded quantum annealing
}
\author{Fernando \surname{Pastawski}}
\affiliation{Institute for Quantum Information and Matter  and Walter Burke Institute for Theoretical Physics, California Institute of Technology, Pasadena, California 91125, USA
}
\author{John Preskill}
\affiliation{Institute for Quantum Information and Matter and Walter Burke Institute for Theoretical Physics, California Institute of Technology, Pasadena, California 91125, USA
}

\begin{abstract}
Recently, Lechner, Hauke and Zoller \cite{Lechner2015} have proposed a quantum annealing architecture, in which a classical spin glass with all-to-all pairwise connectivity is simulated by a spin glass with geometrically local interactions. We interpret this architecture as a classical error-correcting code, which is highly robust against weakly correlated bit-flip noise, and we analyze the code's performance using a belief-propagation decoding algorithm. Our observations may also apply to more general encoding schemes and noise models. 
\end{abstract}

\maketitle

Quantum annealing \cite{Farhi2001} is a method for solving combinatorial optimization problems by using quantum adiabatic evolution to find the ground state of a classical spin glass. 
Hoping to extend the reach of quantum annealing in practical devices,  Lechner {\em et al.} \cite{Lechner2015} have proposed an elegant scheme, using only geometrically local interactions, for simulating a classical spin system with all-to-all pairwise connectivity. 
Their scheme may be viewed as a classical low-density parity-check code (LDPC code) \cite{Galager2003}; here we point out that the error-correcting power of this LDPC code makes the scheme highly robust against weakly correlated bit-flip noise. This observation also applies to other schemes for simulating spin systems based on LDPC codes.

\begin{figure}[t]
\begin{center}
\includegraphics[width=\columnwidth]{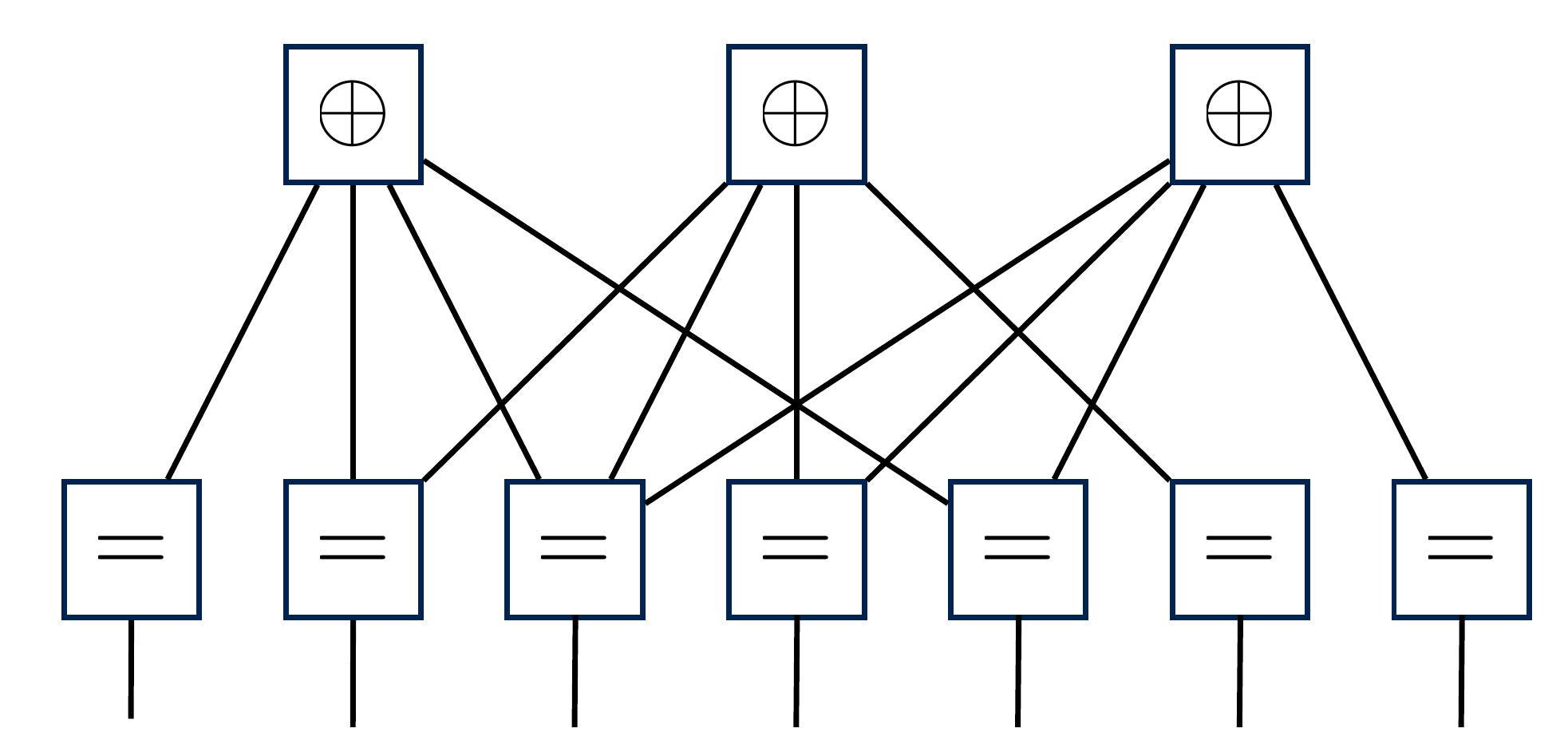}
\caption{The bipartite Forney-style factor graph (FFG) for the $[7,4,3]$ Hamming code, with linear constraint nodes denoted $\oplus$ and variable nodes denoted ($=$). Constraint nodes correspond to rows of the parity check matrix $H$ in eq.(\ref{eq:parity-check}), and variable nodes correspond to columns; an edge connects two nodes if a 1 appears in $H$ where the corresponding row and column meet. 
}\label{fig:LDPCstandard}
\end{center}
\end{figure}

Lechner {\em et al.} propose representing $N$ logical bits $\vec b =\{b_i, i = 1, 2, \dots, N\}$ using $K=\binom{N}{2}$ physical bits $\vec g=\{g_{ij}, 1\le i < j \le N\}$, where $g_{ij}$ encodes $b_i\oplus b_j$ and $\oplus$ denotes addition modulo $2$. 
The $K$ physical variables obey $K-N+1$ independent linear constraints. Hence only $N{-1}$ physical variables are logically independent; we may, for example, choose the independent variables to be $\{g_{12}, g_{23}, g_{34}, \dots , g_{N{-}1,N}\}$. 
The linear constraints may be chosen to be weight-3 parity checks. 
If weight-4 constraints are also allowed then the parity checks can be chosen to be geometrically local in a two-dimensional array. Higher-dimensional versions of the scheme may also be constructed \cite{Lechner2015}; we will discuss only the two-dimensional coding scheme here, but the same ideas also apply in higher dimensions. 

\begin{figure}[t]
\begin{center}
\includegraphics[width=\columnwidth]{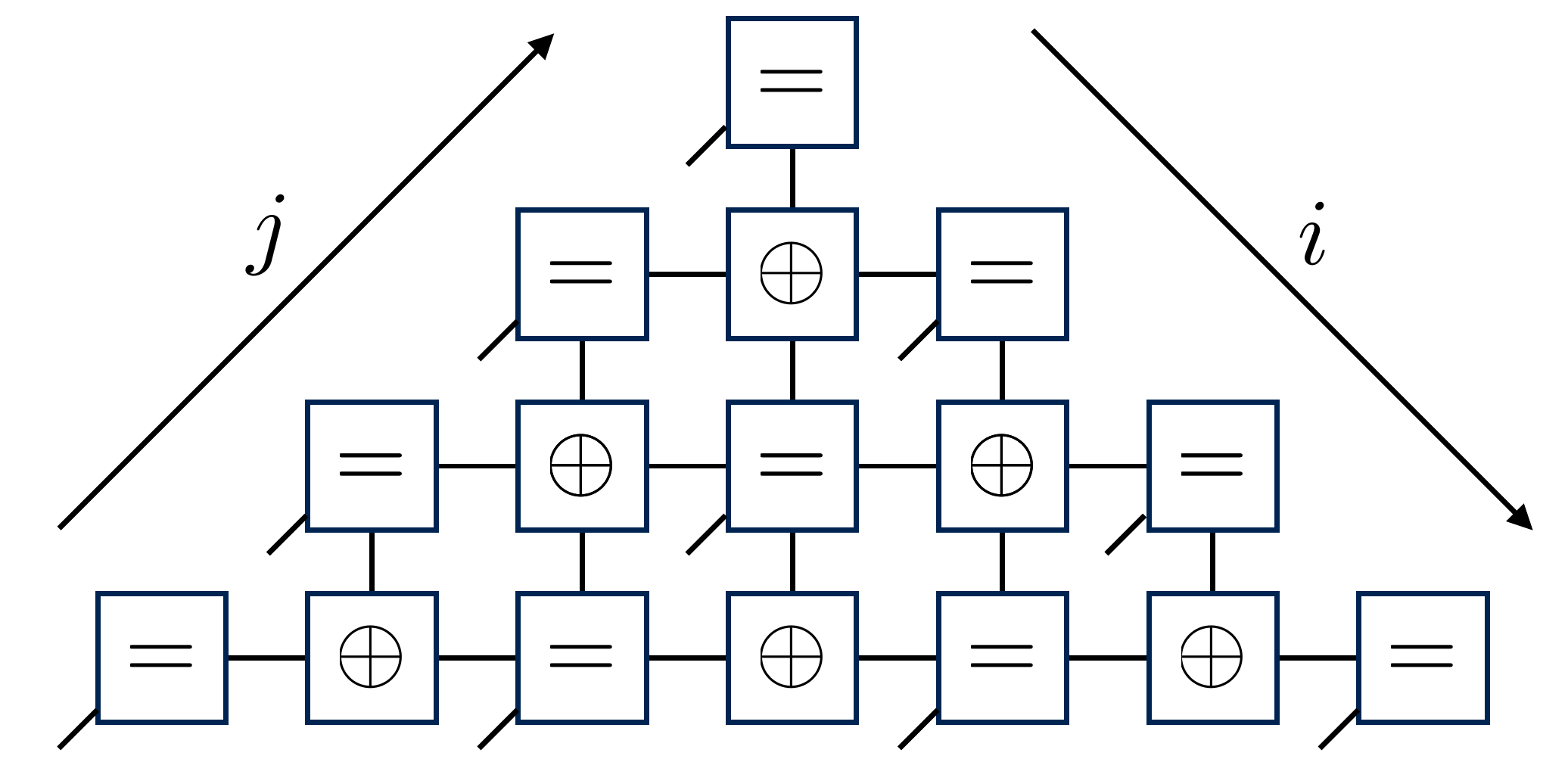}
\caption{
An FFG for the LDPC code of the LHZ scheme, based on Fig.~1d of Ref. \cite{Lechner2015}, shown for $N=5$. Here the variable nodes represent the physical spin variables $\{g_{ij}\}$, with $i\in[1,N-1]$ and $j\in [2,N]$. The geometrically local linear constraints ensure that certain closed loops of spins have even parity. 
}\label{fig:LHZ_FFG}
\end{center}
\end{figure}

An LDPC code is a (classical) linear error-correcting code which can be represented as a sparse bipartite graph called a Forney-style factor graph (FFG), also known as a Tanner graph. To illustrate the FFG concept, Fig.~\ref{fig:LDPCstandard} shows the FFG for the $[7,4,3]$ Hamming code, which has parity check matrix~\footnote{This choice for $H$ matches the FFG in Fig. 1, and also makes manifest the invariance of the code space under cyclic permutations of the bits.}
\begin{align}\label{eq:parity-check}
H = \left(
\begin{matrix}
1&1&1&0&1&0&0\\
0&1&1&1&0&1&0\\
0&0&1&1&1&0&1
\end{matrix}
\right).
\end{align}
The code's parity checks are the linear constraint nodes, denoted $\oplus$ in the graph, while the bits in the code block are the variable nodes, denoted ($=$). All lines connecting to a variable node have the same value (either 0 or 1), and all lines connecting to a constraint node are required to sum to 0 modulo 2. Thus each variable node corresponds to a column of $H$, each constraint node corresponds to row of $H$, and an edge of the FFG connects a variable node and constraint node if and only if $H$ has the entry $1$ in that row and column. The code has ``low density'' in the sense that each parity check has low Hamming weight, and correspondingly each constraint node is connected by edges of the FFG to a small number of variable nodes. The parity check matrix for a particular linear code can be chosen in many ways; hence there are many possible FFG presentations of the same code. Fig.~\ref{fig:LHZ_FFG} shows one possible FFG for the LDPC code of the LHZ scheme. Later we will discuss another FFG for this code.

While $g_{ij}$ denotes the value of $b_i\oplus b_j$ in the ideal ground state of the classical spin glass, we use $g^{\prime}_{ij}$ to denote the (possibly noisy) readout of the corresponding physical variable after a run of the quantum annealing algorithm. 
If the readout is not too noisy, we can exploit the redundancy of the LDPC code to recover the ideal value of $\{b_i\oplus b_j\}$ from the noisy readout $\vec{g}'$ with high success probability. 
Given an error model, we can determine the conditional probability $p(\vec{g}'|\vec b)$ of observing $\vec{g}'$ given $\vec b$. 
Assuming that each $\vec b$ has the same {\em a priori} probability, we decode $\vec{g}'$ by finding the most likely $\vec b$:
\begin{align}
\vec b_{\rm decoded} = {\rm MLE}(\vec{g}')= {\rm ArgMax}_{\vec b}\ p(\vec{g}'|\vec{b}),
\end{align}
where MLE means ``maximum likelihood estimate.'' 
In fact, we can only recover the ideal $\vec b$ up to an overall global flip since one bit of information is already lost during encoding.

We adopt the simplifying assumption of independent and identically distributed (i.i.d.) noise: $g^{\prime}_{ij}$ is flipped from its ideal value $g_{ij}$ with probability $\varepsilon \leq 1/2$, and agrees with its ideal value with probability $1-\varepsilon$.
Though we do not necessarily expect this simple noise model to faithfully describe the errors arising from imperfect quantum annealing, our assumption follows the presentation of \cite{Lechner2015}. This model might be appropriate if, for example, the noise is dominated by measurement errors in the readout of the final state. It also allows us to estimate $p(\vec {g^{\prime}}|\vec b)$, either analytically or numerically. Exact MLE decoding is possible in principle, but has a very high computational cost. We will settle instead for decoding methods which are computationally feasible though not optimal. 

There is a very simple error correction procedure for which we can easily estimate the probability of a decoding error.
For the purpose of decoding (say) $g_{12} \equiv b_1\oplus b_2$, we make use of the following $N{-}2$ weight-3 parity checks:
\begin{align}\label{eq:3termConstraints}
0= (12)\oplus (23)\oplus (13)= \dots =(12)\oplus (2N)\oplus (1N),
\end{align}
where we've used $(ij)$ as a shorthand for $g_{ij}$. 
These checks provide us with $N{-}2$ independent ways to recover the logical value of $b_1\oplus b_2$, namely
\begin{align}
b_1\oplus b_2 = (13)\oplus(23)=(14)\oplus (24)= \dots =(1N)\oplus (2N).
\end{align}
(Of course, $g^{\prime}_{12}$ itself provides another independent way to recover $b_1\oplus b_2$, but to keep our analysis simple we will not make use of $g'_{12}$ here.)
Since $g^{\prime}_{ij}\neq g_{ij}$ with probability $\varepsilon$, each $g^{\prime}_{1j} \oplus g^{\prime}_{2j} \neq g_{ij}$ with probability
\begin{align}
\varepsilon^* := 2 \varepsilon (1 - \varepsilon) \leq 1/2.
\end{align}

Therefore, $g_{12}$ is protected by a length-$(N{-}2)$ classical repetition code with bits flipping independently with probability $\varepsilon^*$. 
The probability of a majority vote decoding error can be estimated from the Chernoff bound:
\begin{align}\label{eq:chernoff}
p_{\rm fail} \le \exp\left(-2(N-2) \left(\frac{1}{2} - \varepsilon^* \right)^2\right).
\end{align}
This is not the tightest possible Chernoff bound, and using additional information such as the observed value of $g^{\prime}_{12}$ will only improve the success probability.
However, eq.(\ref{eq:chernoff}) already illustrates our main point: 
the probability of a decoding error for any $b_i\oplus b_j$ decays exponentially with $N$. 
A simple union bound constrains the probability with which \emph{any} of the $N-1$ bits are decoded incorrectly:
\begin{align}\label{eq:errorDecrease}
p_{\rm fail}^{\rm total} \le (N-1) \exp\left(-2(N-2) \left(\frac{1}{2} - \varepsilon^* \right)^2\right).
\end{align}

Including $g^{\prime}_{12}$ in the decoding algorithm improves the accuracy of our estimate of $b_1\oplus b_2$, and including higher-weight parity checks such as $0=(12)\oplus (23) \oplus (34)\oplus (14)$ can yield further improvements. 
Following a pragmatic approach to using such information, we have implemented \emph{belief propagation} (BP) \cite{Pearl1982}, a fairly standard decoding heuristic for LDPC codes. 
BP efficiently approximates MLE decoding when the constraint graph is a tree, and sometimes works well in cases where the graph contains closed loops. For an introductory account of FFGs and BP see Ref. \cite{Loeliger2004}.

\begin{figure}[!h]
\centering
\begin{varwidth}{0.5\linewidth}  
\subfigure[Prior probabilities]{\includegraphics[width=4cm]{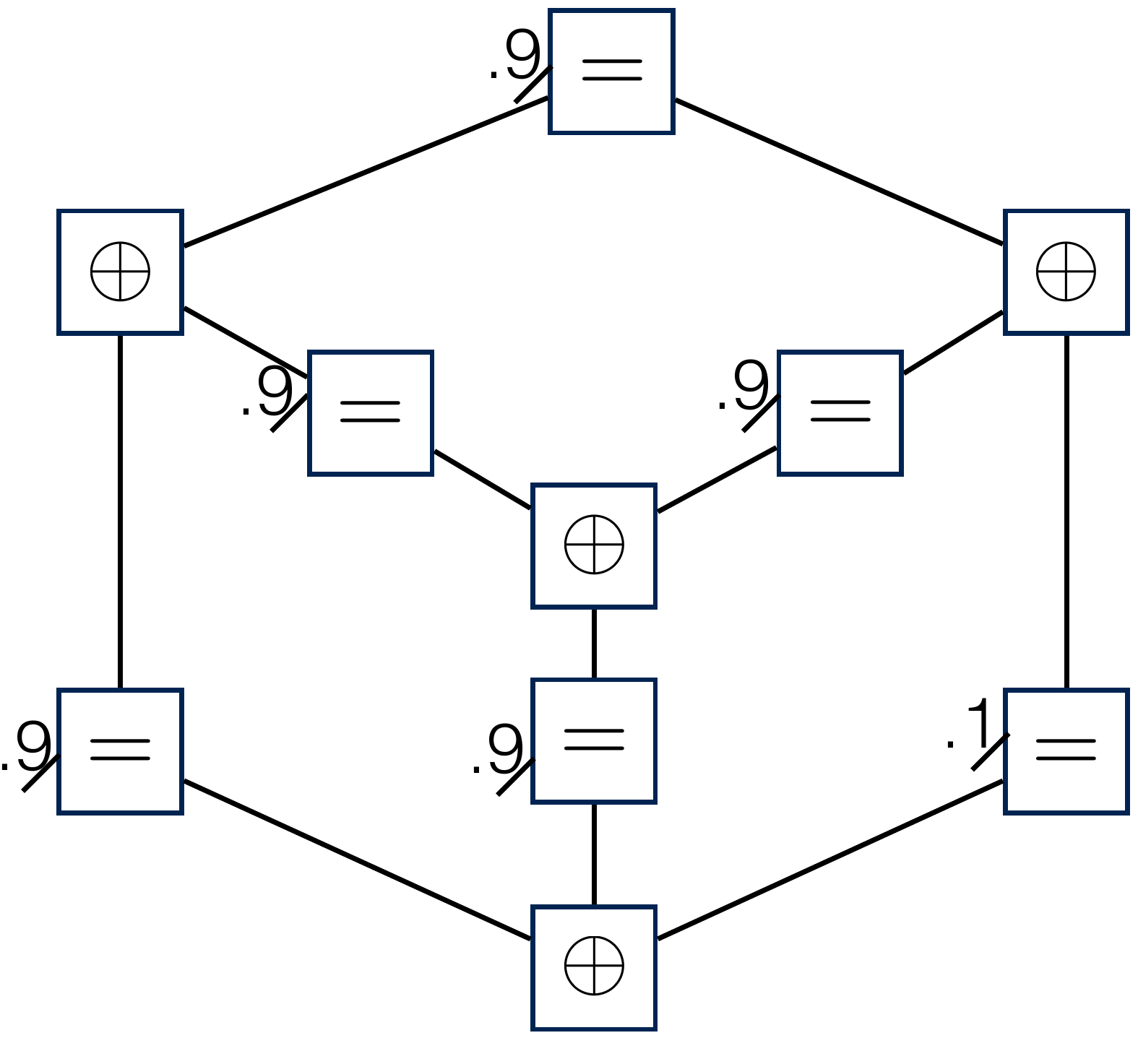}}
\subfigure[Propagation]{\includegraphics[width=4cm]{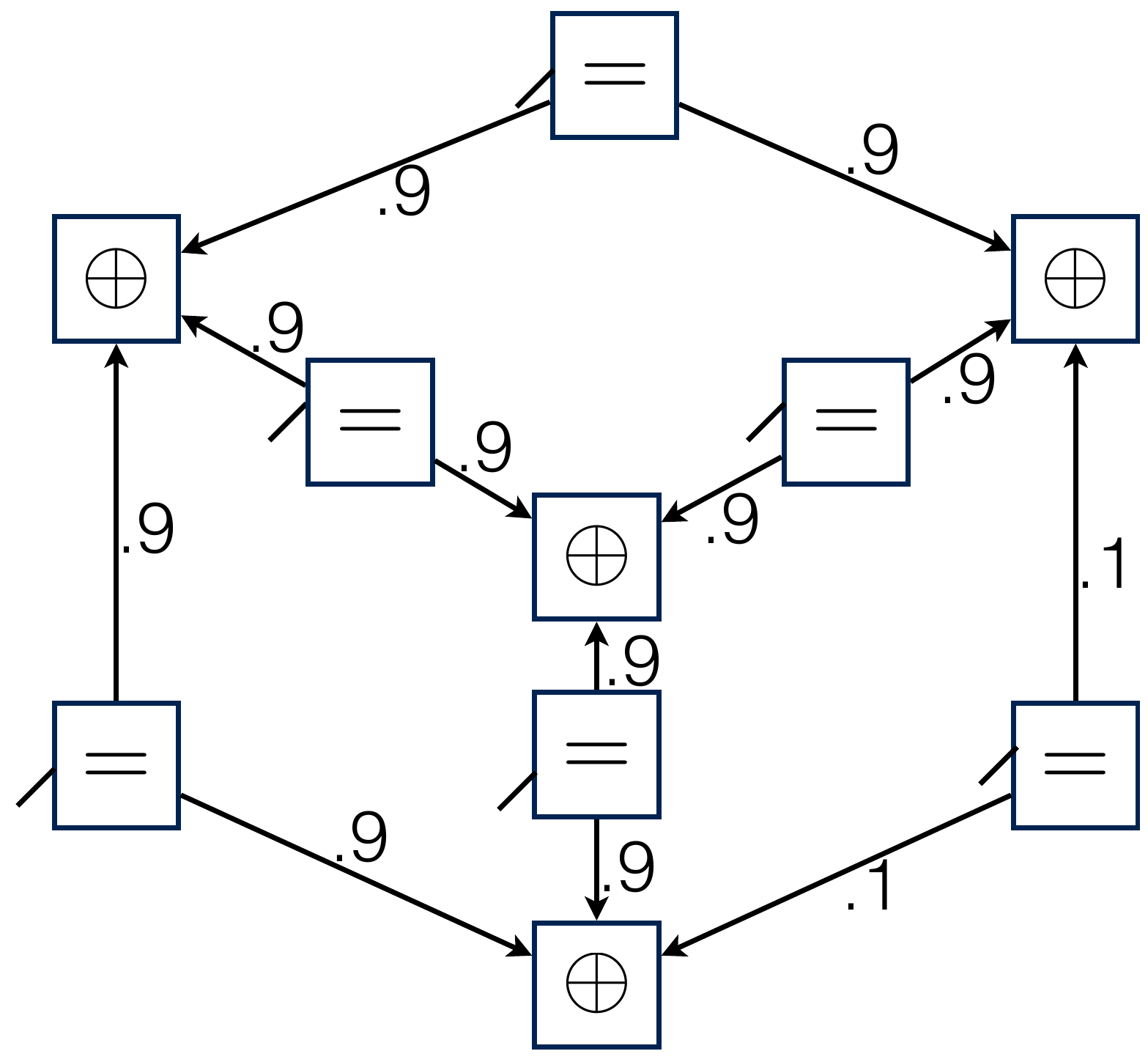}}\\
\end{varwidth}
\begin{varwidth}{0.5\linewidth}  
\subfigure[Sum of products summary]{\includegraphics[width=4cm]{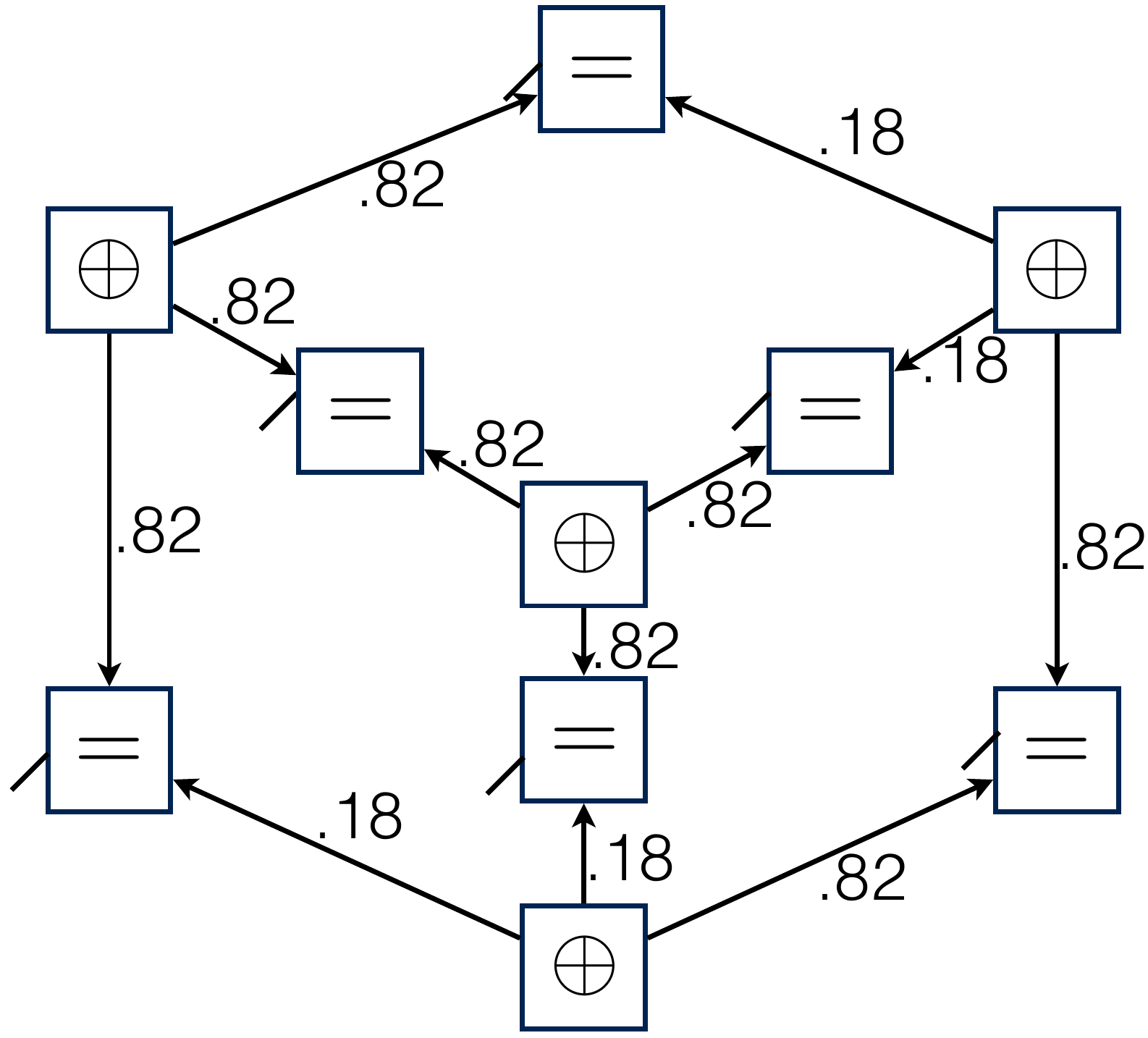}}
\subfigure[Posteriori probabilities]{\includegraphics[width=4cm]{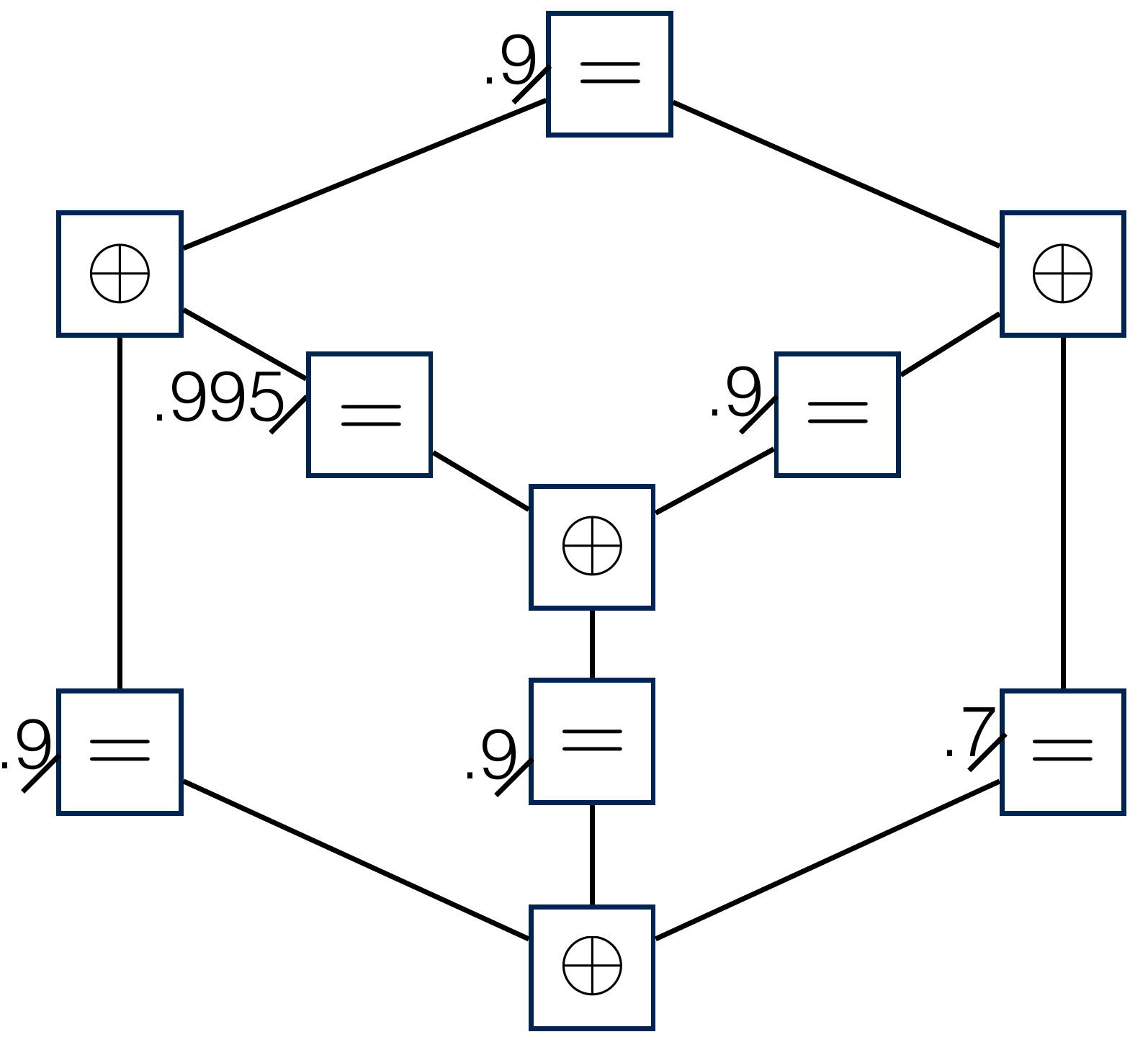}}
\end{varwidth}
\caption{One iteration of the BP realization for the case $\varepsilon=0.1$ and $N=4$. Each number shown is the probability $g_{ij}(0)$ that the associated node $(ij)$ has the value $0$. 
(a) The prior distribution assuming each measured physical spin has the value 0, except for one spin in the lower right corner which has value 1. This value is incompatible with the rest, indicating a likely error.
(b) Values of $g_{jk}(0)$ passed from variable nodes $(=)$ to neighboring constraint nodes $\oplus$.
(c) Values of $\left(g_{ij}\right)_{a\to v}$, computed by the sum-product formula, passed from constraint nodes to neighboring variable nodes.
(d) Updated {\em a posteriori} values for $g_{ij}(0)$, calculated as the (normalized) product of received messages and prior probabilities.
} \label{fig:BPillustration}
\end{figure}

In BP, a marginal distribution is assigned to each variable, and updated during each iteration based on the values of neighboring variables on the FFG. Therefore, the implementation of BP depends not only on the code and the noise model, but also on how the code is represented by the FFG. For our implementation, rather than using the FFG in Fig.~\ref{fig:LHZ_FFG}, with $\binom{N-1}{2} = O(N^2)$ constraint nodes, we use an FFG with $\binom{N}{3}=O(N^3)$ constraint nodes instead. For each triplet $(b_i,b_j,b_k)$ of logical bits, the corresponding constraint is 
\begin{align}
0= (ij)\oplus (jk)\oplus (ik)
\end{align}
in the notation of eq.(\ref{eq:3termConstraints}). These constraints are highly redundant, and the larger number of constraints increases the cost of each BP iteration. On the other hand, this scheme has the advantage that it treats all variables symmetrically, and furthermore it includes all the constraints used in our majority voting scheme, which we have already seen has a noise threshold of $1/2$ for i.i.d.~noise in the limit of large $N$, ensuring that BP will also converge to the correct answer in this limit. Our FFG is shown in Fig.~\ref{fig:BPillustration} for $N=4$, in which case the FFG is planar, with six variable nodes and four constraint nodes. For large values of $N$ the FFG is highly connected and hard to draw. 

In a single iteration of BP, the marginal probability distributions assigned to the variables are updated by the following two-step procedure. In the first step, each constraint node sends a message to each of its neighboring variable nodes. For the edge of the FFG connecting constraint node $a$ to variable node $v$, this message, computed using the sum-product formula, is constraint node $a$'s guess regarding the marginal distribution for $v$, based on the marginal distributions for its other neighbors besides $v$. To be concrete, in the FFG for the LHZ code, let $g_{ij}(0)$ denote the probability that variable $g_{ij}$ has the value $0$, and let $g_{ij}(1)$ denote the probability that $g_{ij}=1$. The message sent by the constraint node $a=(12)\oplus(23)\oplus(13)$ to the variable node $v =(12)$ is 
\begin{align}
  \binom{g_{12}(0)}{ g_{12}(1)}_{a\to v} = \binom{g_{23}(0)g_{13}(0)+g_{23}(1)g_{13}(1)}{g_{23}(0)g_{13}(1)+g_{23}(1)g_{13}(0)},
\end{align}
where $\left( g_{12}\right)_{a\to v}$ denotes $a$'s guess. In the second step of the procedure, each variable node updates its marginal distribution by evaluating the normalized product of its previous {\em a priori} probability and all estimated probabilities passed by the neighboring constraint nodes. To be concrete, suppose that variable node $v$ is connected by edges to constraint nodes $a$ and $b$; then the updated probability distribution for variable node $v$ will be 
\begin{align}
  \binom{g_v(0)}{ g_v(1)}_{\rm updated}\propto \binom{g_v(0) g_{a\to v}(0) g_{b\to v }(0)}{g_v(1) g_{a\to v}(1) g_{b\to v }(1)},
\end{align}
up to normalization. 

For an i.i.d.~noise model with error probability $\varepsilon$, we assign initial distributions to each variable node by assuming that the observed value of $g_{ij}$ is correct with probability $1-\varepsilon$ and incorrect with probability $\varepsilon$. To decode, probabilities are updated repeatedly until they converge to stable values or until the decoding runtime has elapsed. Intuitively, a consistent neighborhood reduces the entropy of the local marginal distributions, whereas an inconsistent neighborhood may increase the entropy or even change a variable's most likely value. How inconsistencies are resolved is illustrated in Fig.~\ref{fig:BPillustration}, which depicts one iteration of BP for the LHZ code with $N=4$. There, the marginal distribution of one variable node is incompatible with the rest, and its updated distribution favors a flipped value, correcting the error.


\begin{figure}[t]
\begin{center}
\includegraphics[width=\columnwidth]{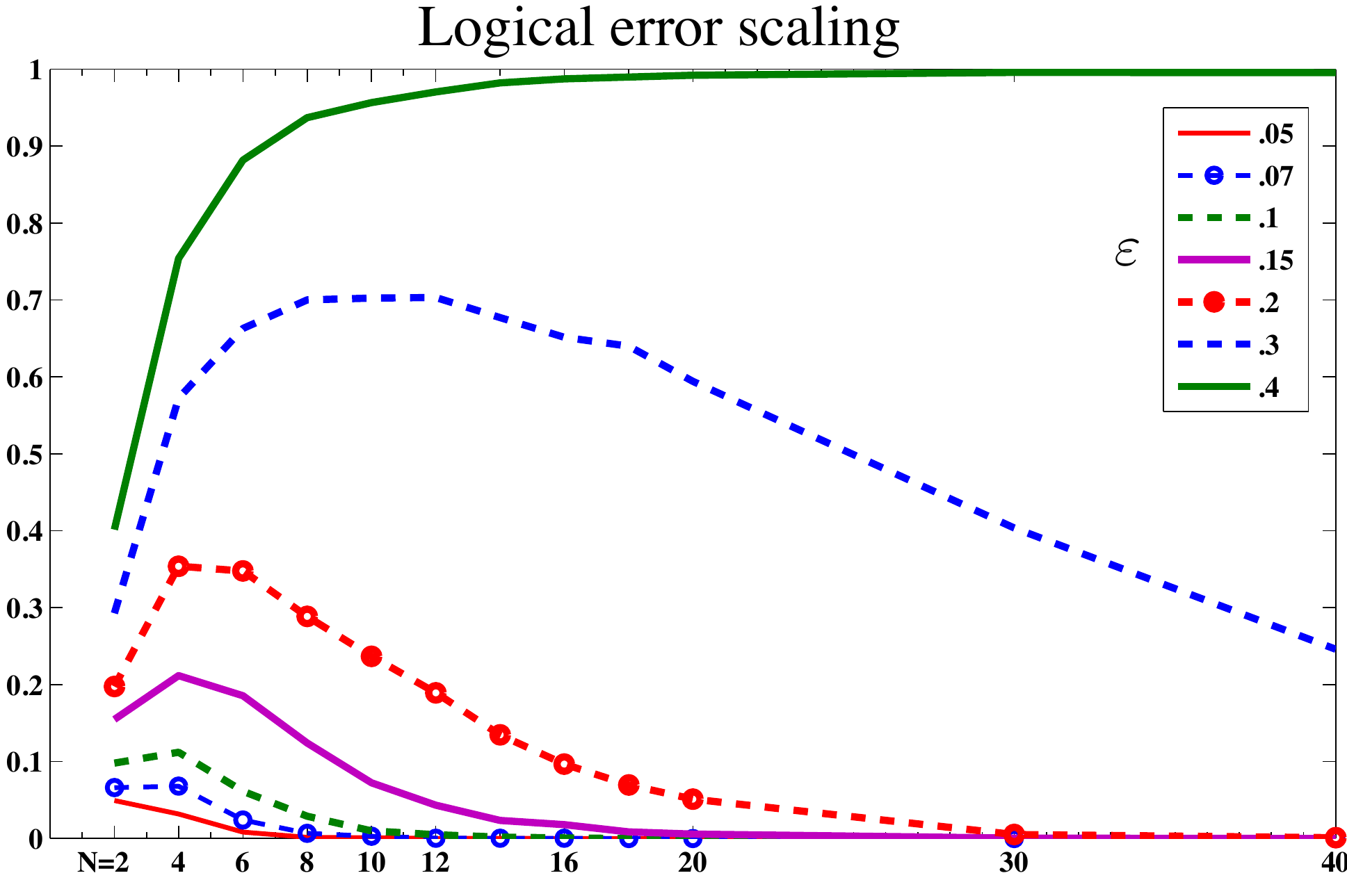}
\caption{
Performance of iterative BP decoding algorithm. 
The probability of a decoding error is plotted as a function of the number $N$ of encoded spins, for various values of the physical error probability $\varepsilon$. Each data point was obtained by averaging over 5000 noise realizations, and for each realization the BP algorithm was iterated five times, incorporating information about loops up to length $33=2^5+1$. The decoding performance is significantly better than for a single BP iteration, where only loops of size $\le 3$ are considered. 
The logical error probability starts at $p_{\rm fail}^{\rm total}= \varepsilon$ for $N=2$ and rises with $N$ until the onset of exponential decay, which begins for a smaller value of $N$ than suggested by eq.(\ref{eq:errorDecrease}).
}\label{fig:BPperformance}
\end{center}
\end{figure}

For the LHZ code and i.i.d.~noise the numerically estimated probability $p_{\rm fail}^{\rm total}$ of a decoding error is plotted in Fig.~\ref{fig:BPperformance} as a function of the error probability $\varepsilon$ and the number $N$ of encoded spins, for $N$ ranging from 2 to 40.  As expected, we find that the failure probability falls steeply as $N$ increases if $\varepsilon$ is not too close to the threshold value $1/2$. Also as expected, $p_{\rm fail}^{\rm total}$ is substantially smaller than the crude estimate in eq.(\ref{eq:errorDecrease}).

We conclude that the architecture proposed in \cite{Lechner2015}, and the decoding method proposed here, provide good protection against i.i.d.~noise in the readout of the physical spins, assuming an error probability $\varepsilon$ for each physical spin  which is independent of the total number $N$ of encoded spins. More generally, we expect powerful decoding strategies such as BP to enhance the performance of other quantum annealing schemes in which the simulated spins are the logical bits of an LDPC code. We note that BP and other related methods have also been used to solve combinatorial problems in a purely classical setting \cite{Braunstein2005}. Perhaps sophisticated classical decoding strategies and quantum annealing, when used together, can solve problems which are beyond the reach of either method used alone. 



To keep our analysis simple, we assumed an i.i.d.~noise model for the physical spins, which might not be an accurate description of the noise in realistic quantum annealing. In fact, Albash {\em et al.} \cite{Albash2016} have recently provided evidence that this noise model is inadequate, by investigating the performance of the LHZ scheme using simulated quantum annealing, a Monte Carlo method (using a classical computer) for approximating the behavior of a quantum annealing procedure. The output distributions in actual quantum annealing experiments have been found to agree reasonably well with simulated quantum annealing predictions, and the numerical results in Ref.~\cite{Albash2016} indicate that the LHZ scheme does not outperform the annealing architectures used in current experiments \cite{Bunyk2014}, even after including a final decoding step.
Perhaps quantum error-correcting codes can be invoked to achieve further improvements in performance \cite{Jordan2006, Lidar2008}, but so far no truly scalable scheme for quantum annealing has been proposed \cite{Young2013}. 
How well the Lechner {\em et al.} architecture performs under realistic laboratory conditions is a question best addressed by experiments.

We thank W. Lechner and E. Crosson 
for useful comments and discussions.
FP and JP gratefully acknowledge funding provided by the Institute for Quantum Information and Matter, a NSF Physics Frontiers Center with support of the Gordon and Betty Moore Foundation, and by the Army Research Office.

\end{document}